\documentclass{iaus}
\usepackage{graphicx}

\def\msol{ {$\mathcal M_{\odot}$} }

\def\bdmath{\begin{displaymath}}
\def\edmath{\end{displaymath}}

\def\apj{ApJ}
\def\apjl{ApJL}

\def\aap{A\&A}

\def\avg#1{\overline{\langle #1\rangle}}

\title[Massive Star Hydrodynamics]{Hydrodynamic Processes in Massive Stars}

\author[Meakin]{Casey A. Meakin$^{1,2,3}$}
\affiliation{
  $^1$Astronomy Department, University of Arizona, Tucson, AZ 85721, USA\\
  [\affilskip]
  $^2$Astronomy and Astrophysics Center, University of Chicago,Chicago, IL 60637, USA\\
  [\affilskip]
  $^3$Joint Institute for Nuclear Astrophysics, University of Chicago, Chicago, IL, 60637, USA
  \break email: casey.meakin@gmail.com
}

\pubyear{2008}
\volume{xxx}
\pagerange{000--000}
\date{?? and in revised form ??}
\jname{Proceedings Title IAU Symposium}
\editors{A.C. Editor, B.D. Editor \& C.E. Editor, eds.}
\begin{document}

\maketitle

\begin{abstract}
  The hydrodynamic processes operating within stellar interiors are far richer
  than represented by the best stellar evolution model available.
  Although it is now widely understood, through astrophysical simulation and relevant 
  terrestrial experiment, that many of the basic assumptions which underlie 
  our treatments of stellar evolution are flawed, we lack a suitable, comprehensive 
  replacement.  This is due to a deficiency in our fundamental
  understanding of the transport and mixing properties of a turbulent, reactive, magnetized 
  plasma; a deficiency in knowledge which stems from the richness and variety of
  solutions which characterize the inherently non-linear set of governing equations.
  The exponential increase in availability of computing resources, however, is ushering in
  a new era of understanding complex hydrodynamic flows; and although this field is still 
  in its formative stages, the sophistication already achieved is leading to a dramatic
  paradigm shift in how we model astrophysical fluid dynamics.
  We highlight here some recent results from a series of multi-dimensional stellar interior 
  calculations which are part of a program designed to improve our one-dimensional treatment of 
  massive star evolution and stellar evolution in general.
\end{abstract}

\firstsection 

\section{The Challenge at Hand}

\par Massive stars play a central role in a variety of astrophysical contexts:

\begin{enumerate}
\item Nucleosynthetic yields and galactic chemical evolution
\item Black hole and neutron star formation rates; supernova and gamma ray burst rates
\item ``feedback'' with the ISM and IGM through winds, ionizing photons, and explosions
\item Supernova theory through progenitor evolution and global asymmetries due to convection
  and rotation
\end{enumerate}

\par Each of these diverse topics depends deeply on our ability to 
correctly model the life and death of an individual massive star.
But the evolution of a massive star relies critically on the correct treatment of 
the hydrodynamic transport processes which are operating throughout the
stellar plasma. The difficulty encountered in modeling the same
transport properties even in the far less exotic conditions present in the Earth's 
atmosphere and oceans is a stark reminder of the challenges faced by the
stellar evolutionist.

\subsection{Moore's Law}

{\em\large "Make no little plans. They have no magic to stir men's blood and probably 
  will not themselves be realized."} 
{\bf ---Daniel Burnham, American architect and urban planner}.

\par While computational tools and numerical experiments are beginning to 
provide profound new insights into the nature of turbulent flows, the astrophysical
conditions encountered in stellar interiors reminds us that we have a long way
to go before we can fully resolve such flows.

\par For instance, consider the ratio between the largest length scale present in a stellar
convection zone and the smallest scales on which velocity fluctuations can occur before
being smoother out by viscous forces. In a turbulent medium the smallest scales are
connected to the largest scales through a cascade of energy.  The ratio between largest
and smallest scales can be related to the Reynolds number of the flow if we adopt
the Kolmogorov energy spectrum (\cite{kolmogorov1941,kolmogorov1962}), 
so that $l_{max}/l_{min}\sim {\rm Re}^{3/4}$ (see \cite{boris2007}).
An often cited example is that for the conditions present in the solar convection zone, where
we find something like ${\rm Re}\sim10^{10}$, so that $l_{max}/l_{min}\sim 3\times 10^7$.
Therefore, to model a cubical region which contains all of the relevant scales from the 
largest eddy to the viscous damping scale we would required our calculation to contain
along the lines of $N\sim10^{22}$ computational cells.  For comparison, the largest turbulence
calculations carried out to date have $N = (2048)^3 \sim 10^{10}$ (e.g., \cite{kritsuk2007}) 
to $N=(4096)^3\sim0.7\times 10^{11}$ (on the Earth Simulator)
computational cells. Therefore, we need an increase in computing resources by a 
factor of $\sim 10^{12}$.

\par Moore's law states that the computational resources available for a given cost
doubles every 18 months,

\begin{equation}
  \log_2({\rm flops}/\$) = {\rm time} / (18 {\rm months}).
\end{equation}

\noindent  How long we must wait until a fully resolved simulation of stellar
turbulence is possible at a funding level comparable to current computational astrophysics levels?
If Moore's law holds, we will be able to afford a computing cluster which is a faster by a
factor of $10^{12}$ in  $t \approx 18 {\rm months} \times \log_2 10^{12} \approx 60$ 
years.\footnotemark\footnotetext{Some of the implications of management strategies when
considering such large calculations in the context of Moore's law are examined 
in \cite{gottbrath1999}.}

\par Computers have just surpassed the petaFLOPS barrier, performing
one thousand trillion ($10^{15}$) floating point operations per second (FLOPS).  At this
computing speed, it would take $\sim$4 months to compute just one floating point operation 
per cell in our fully resolved $10^{22}$ zone stellar turbulence calculation.  Although 
this remains a prohibitively expensive calculation, it is exciting that a hundred fold 
increase in speed has been achieved since 2002 (when the 10 terraFLOPS mark was passed), 
just 6 years ago\footnotemark
\footnotetext{See http://www.top500.org for a summary of the fastest computing systems
  in operation as well as historical data.}. (This is equivalent to a Moore's law
doubling period of only $\sim$11 months.) This example illustrates the
telescoping nature of technological advance summarized by Moore's law, and provides a 
truly visceral sense of realism about our earlier estimate that a fully resolved stellar 
turbulence calculation will be feasible in only 60 years.

\par Do these considerations lead us to the conclusion that it is premature to perform 
stellar convection simulations, and suggest that we should instead wait until adequate 
computational resources are available? There are several grounds on which to reject 
this line of reasoning. (1) Significant development is needed in software and data management 
strategies which is arguably best approached by pushing our present resources to their limits.
(2) Analyzing and designing numerical experiments is also a developing art, and
we are still learning how to query data in order to inspire and test new theoretical 
ideas; a creative process which is also best approached by getting our hands dirty.
(3) It is possible that many of the resulting flow features captured by our incompletely
resolved numerical experiments are nevertheless robust because of an inherently universal
property of turbulent flow; the ``turbulent cascade''.  We briefly discuss this topic
in the next subsection.

\subsection{The ILES Approach}

\par The large eddy simulation (LES) approach entails explicitly modeling the largest
eddies in a turbulent flow and using a turbulence model to incorporate the
mixing, dissipation, and dynamical consequences of the smaller, unresolved scales.
One of the motivating factors behind this approach is that the majority of the
kinetic energy in the flow is contained in the largest scales.  Another, relates
to how the largest and smallest scales of motion couple in a turbulent flow
through an inertial range cascade which. As observed by \cite{boris2007}:
``The physically important aspects of the fluid dynamics of turbulent flow can be
notably insensitive to the small-scale details of how it is computed.''

\par This latter observation underlies a somewhat recent shift in perspective about 
how to model turbulence and carry out LES simulations.  In particular, there has
been a shift away from developing sophisticated subgrid turbulence models, and
instead taking advantage of the insensitively of the large scale motions on the
detailed properties of the smallest scale motions where dissipation occurs.
Instead, the basic physically motivated numerical algorithms used in 
modern hydrodynamics methods ensure that (1) conservation, (2) monotonicity, 
(3) causality, and (3) locality are built into the solutions (\cite{boris2007}).
This approach has been dubbed Implicit Large Eddy Simulation (ILES).

\section{New Resources, New Tools}

\par A number of groups have begun to model stellar interiors and atmospheres
in multi-dimensions using simulation codes designed for massively parallel processing 
environments; platforms which employ thousands to hundreds of thousands of microprocessors 
simultaneously on a single calculation.  Developing the software infrastructure necessary to
perform numerical simulations on modern equipment is just as important as the advances
in hardware manufacturing which Moore's law describes, particularly in light of the changing
face of parallel processing architectures.  Multi-core processors (cell processors) 
and hardware heterogeneity is likely to play a prominent role in the 
future (\cite{turner2007}).  These changes in computing architecture depend on advances
in multi-threading programming capabilities.
Software and information management technologies are also needed to effectively manipulate
data sets which will soon exceed a petabyte (1 PB = 10$^3$ TB = $10^6$ GB).  Post processing
data is already beginning to use a significant fraction of the total number of FLOPS
required for a computational project.

\par Vast parallelism favors numerical schemes which have a high degree of communication 
locality. Anelastic (\cite{gough1969}) and implicit methods (including low-Mach number 
solvers e.g. \cite{almgren2006,lin2006}) require solutions to elliptic equations which are 
burdened by {\em global} communication at each time step, an operation which is not ideally 
suited to large parallel platforms. The computational advantages that these approximate 
methods have traditionally held over explicit schemes are being 
lost in the new era of massively parallel computing. The good news is that the enormous
increases in computing resources is making fully compressible, multi-physics, explicit
solutions accessible for an increasingly rich set of astrophysical problems.
Modelers are now able to incorporate a higher level of realism into their models, including
magnetic fields, realistic equations of state, sophisticated radiation transport schemes, 
multi-species flow, and combustion physics (i.e., nuclear reactions).
While some concerns have been raised concerning the use of fully compressible solvers
for low Mach number flows (\cite{schneider1999}), it isn't clear that these short comings are 
afflicting present simulations of stellar convection. Direct comparison between
fully compressible, anelastic methods, and analytic results for very low mach number
flow ($M < 10^{-3}$; see \cite{meakin2007b}) suggest a promising outlook for 
fully compressible solvers such as the piecewise parabolic method (PPM) of \cite{colella1984}.

\section{New Data}

\par The nuclear timescale $\tau_{nuc} = X/\dot{X}$, for fuel $X$ which characterize
the different evolutionary phases in a stars life are generally many orders of
magnitude larger than the advective timescale $\tau_{adv} = L/v$, for fluid with
a speed $v$ traversing a region of size $L$. These disparate timescales make
computing the entirety of a stars life in three dimensions prohibitively expensive
in the foreseeable future. However, the condition $\tau_{nuc}\gg\tau_{adv}$ allows us 
to separate the problem so that we can study a snapshot of the evolution, and use this
snapshot to formulate a theory of stellar hydrodynamics which we can then incorporate
into a 1D stellar evolution code.  During the later burning phases
and just prior to core collapse in massive stars, we find $\tau_{nuc}\sim\tau_{adv}$.
Under these circumstances, the timescale for nuclear evolution becomes small enough
to simulate directly.

\subsection{Pre Core-collapse Silicon Shell Burning and Symmetry Breaking}

\par
\begin{figure}
  \includegraphics[scale=0.26]{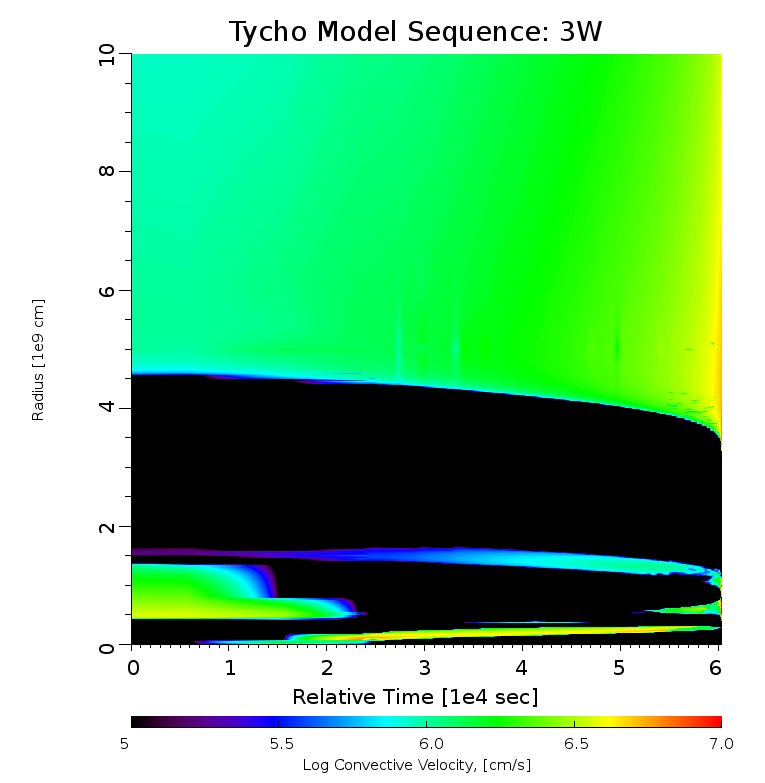}%{f1a.ps}%{late.cv.ps}
  \includegraphics[scale=0.37]{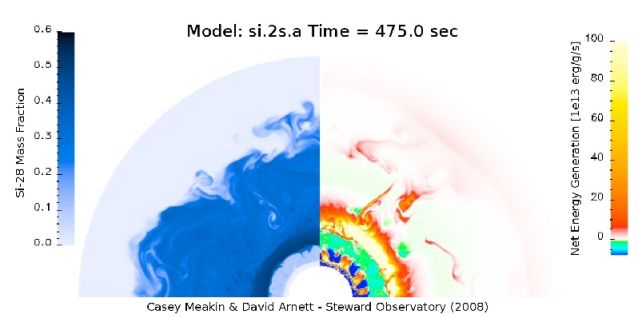}%siburn.ps}
\caption{Pre-supernova silicon burning hydrodynamics: (left) The radial dependence of
the convective velocity is shown as a function of time for a one-dimensional 23\msol
stellar model as it approaches core collapse, which commences at the very end of the
time-sequence shown. The innermost convection zone is due to silicon burning, and a
transition from core to shell burning can be seen.  The overlying convection zones
are driven by oxygen, neon, and carbon burning shells. This model was evolved with
the TYCHO stellar evolution code. (right) This snapshot shows the distribution of
$^{28}$Si and net energy generation for a two dimensional hydrodynamic simulation of
the TYCHO model $\sim$1000 s before core collapse. Silicon, oxygen, neon, and carbon
are burning in concentric shells progressively further away from the iron-rich core
which will soon undergo gravitational collapse. The outer boundary of the oxygen
burning convection zone is strongly perturbed by the convective motions which
eventually mixes the carbon, neon and oxygen burning shells together prior to core
collapse (Meakin 2006b).\label{fig:silicon-burning}}
\end{figure}

 We have begun a program of multi-dimensional stellar interior modeling which
tackles both the quasi-steady and dynamic evolution.  Some preliminary work on simulating the
reactive hydrodynamic flow associated with pre-core collapse silicon burning in a shell 
which surrounds an iron core is described in \cite{meakin2006} and \cite{meakin2007a}
(see Figure~\ref{fig:silicon-burning}). An interesting discovery is the strong interaction
between the turbulent convection and the intervening stably stratified layers.  Stable
layers are significantly distorted by the convective motions, allow for coupling between
different burning zones through waves excited in the stable layers (wave cavities), and
significant amounts of material is entrained from the convective boundaries into the
burning zones.  These effects lead to large asymmetries as core collapse is approached
which could play an important role in seeding instabilities and affecting the outcome
of core collapse and the subsequent supernova explosion.

\subsection{Quasi-Steady Oxygen Shell Burning}
\par
\begin{figure}
\begin{center}
\includegraphics[scale=1.4]{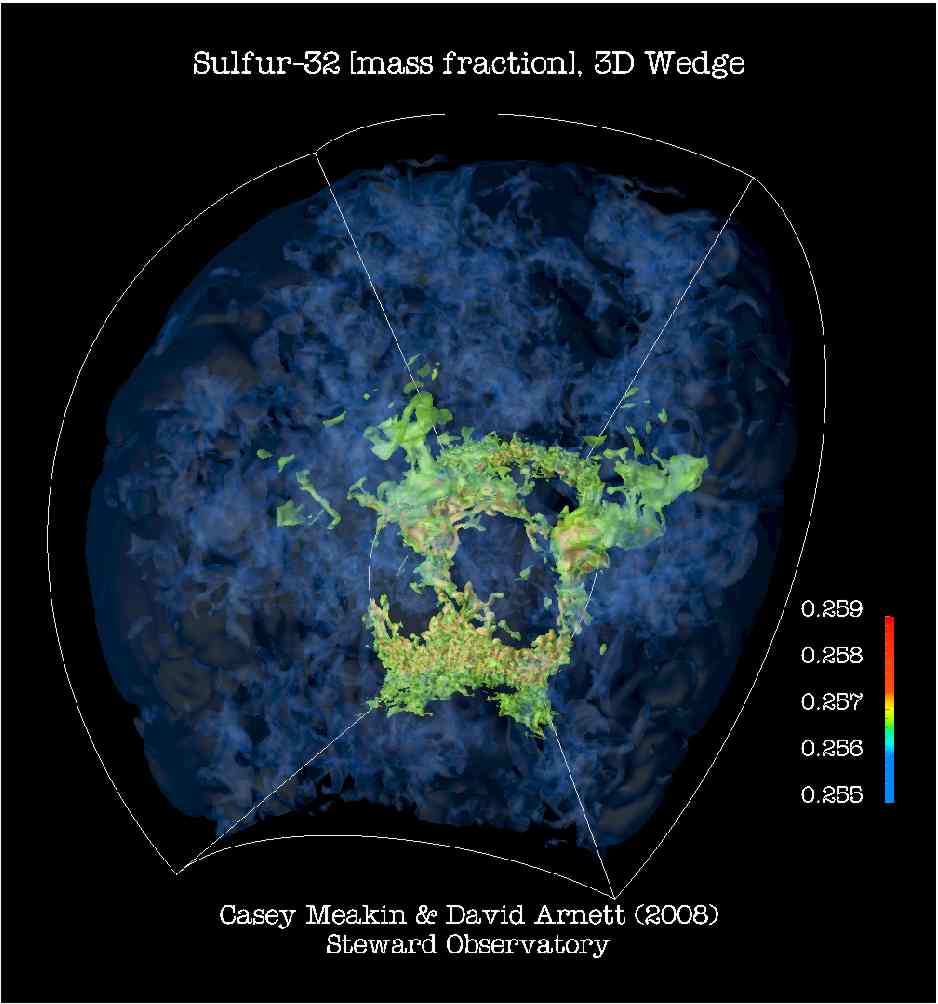}%{ob.4hp.m.ps}
\end{center}
\caption{The turbulent flow field in a deep (4 pressure scale height) oxygen burning shell
  is shown for a computation with a large angular domain (120$^{\circ}\times120^{\circ}$),
  the boundaries of which are shown outlined by white lines.
  The domain, which is described by spherical polar coordinates, is oriented so that the
  polar direction is roughly in the up-down direction, and the azimuthal direction is
  oriented roughly in the left-right direction.
  The mass fraction of $^{32}$S is visualized in order to give a sense of the topology
  and the complex, multi-scale, turbulent nature of the flow.
  Material with a high mass fraction of $^{32}$S is being entrained into the turbulent
  oxygen burning convective shell from the underlying silicon and sulfur rich core.  
  The computational domain contains 17 million cells.  Evolving the flow
  for 5 convective turnover times requires $\sim$1 million cpu-hours on a computing 
  cluster equipped with quad Intel Xeon EM64T 2.8GHz processor cores.
  (Data from Meakin \& Arnett, 2008 in preparation.
  \label{fig:oxygen-burning})
}
\end{figure}

 The neutrino cooled oxygen shell burning epoch is an ideal evolutionary phase 
to study the physics of quasi-steady state stellar convection.  The acceleration 
of this burning stage due to neutrino cooling reduces the ratio between 
the thermal and hydrodynamic timescales, hence easing the burden of obtaining a
relaxed model (see e.g. \cite{arnett1996}, Ch. 10).  Recently, we have extended oxygen 
shell burning simulations to include significantly larger computational domains, longer
evolutionary timescales, and 3D flow (\cite{meakin2006b,meakin2007b,meakin2007c}).
A snapshot of the turbulent flow within an oxygen burning shell is presented in 
Figure~\ref{fig:oxygen-burning}.  While we find a statistically converged, smooth,
quasi-steady state, it is characterized on smaller timescales by significant intermittency
and fluctuations.  This is illustrated in Figure~\ref{fig:buoyancy-work}, in which we
present both a time averaged radial profile and a space-time diagram showing the 
evolution of the buoyancy work in a convective oxygen burning shell.

\section{Processes and Theory}

\par It is important to bear in mind that numerical simulations of stellar convection
are not complete and faithful representations of the actual flows present within a stellar 
interior.  What simulation does provide is a fully non-linear solution with a
large number of degrees of freedom which is constrained by an ever more realistic 
astrophysical context (equation of state, background structure and source terms, 
better nuclear energetics, etc).
These solutions provide the theorist with (1) insight into the fundamental processes
which might be operating in a stellar interior, and (2) estimates for the amplitudes
and length scales present in the flow which drive instabilities on smaller,
unresolved scales.  The data from these numerical experiments which inspire theoretical 
ideas, must ultimately be augmented by a richer, and broader theory of basic 
processes.

\subsection{Reynolds Averaged Equations}

\par We develop a kinetic energy (KE) equation in \cite{meakin2007c} by
decomposing the velocity $\mathbf{u}$, density $\rho$, and pressure $p$ fields into mean and 
fluctuating components $\varphi = \varphi_0 + \varphi'$, employing
the hydrostatic equilibrium condition, and performing averages,

\begin{equation}\label{tke-eq}
  \partial_t \avg{\rho E_K} + \nabla\cdot\avg{\rho E_K\mathbf{u_0}} =
  -\nabla\cdot\avg{\mathbf{F_p} + \mathbf{F_K}} +
  \avg{p'\nabla\cdot\mathbf{u'}} + \avg{\mathbf{W_b}} -\varepsilon_K
\end{equation}

\noindent where $E_K$ is the kinetic energy per gram, $\mathbf{W_b}$ is the buoyancy
work term,  $\varepsilon_K$ is the viscous dissipation of kinetic
energy, $p'\nabla\cdot\mathbf{u'}$ represents the
compressional work done by turbulent fluctuations, and $\mathbf{F_K}$ and $\mathbf{F_p}$ 
are kinetic energy and pressure-correlation fluxes.  A complimentary equation for the internal 
energy can may be developed (see \cite{meakin2007c,arnett2008}).

\par One of the primary aims in turbulence (and stellar convection) research
is to develop physical models for 
the various terms in these equations, such as the dissipation and flux terms.  The
reliability of these model terms is only as good as the physical assumptions
on which they are based.  Often, one is forced to resort to mathematically motivated,
ad hoc or phenomenologically based closure models to develop a working theory, and these
``theories'' are often replete with adjustable parameters which absorb our ignorance
about various flow properties.  For instance, a commonly used model
for the kinetic energy flux is to assume (e.g., \cite{stellingwerf1982,kuhfuss1986}),

\begin{equation}
  \mathbf{F_K} \propto -\nabla(E_K)
\end{equation}

\noindent which is sometimes referred to as the down gradient approximation (DGA).  
Although this model
is contradicted by experiment, fundamental theory, and numerical simulation 
(see e.g., \cite{pope2000}) it remains the cornerstone of many modern 
turbulence theories which are used in stellar evolution modeling.  

\par One avenue for moving beyond simplified turbulence models and closures such as 
DGA is to draw upon the physical intuition garnered from (1) ever 
more realistic numerical simulation, (2) better laboratory flow visualization techniques, 
and (3) cross pollination between the different fluid dynamics sub-discplines 
(e.g., oceanography, astrophysics, laboratory combustion, etc).

\subsection{Dissipation and the Mixing Length}

\par The dissipation of the piecewise parabolic method (PPM) acts on the
smallest scales. The numerical dissipation characteristics of this 
hydrodynamics algorithm compares remarkably well to other approaches used to model 
turbulence (see \cite{benzi2007,boris2007}).  
The good energy conservation properties
of the finite volume, conservative PPM scheme allows us to infer the dissipation
rate of kinetic energy in our convection simulations, which is shown in 
Figure~\ref{fig:oxygen-analysis}(left). Guided by the dependence of dissipation on
the kinetic energy scale of the flow in homogeneous turbulence, we posit
that the dissipation in the convective shell can be written as,

\begin{equation}
  \varepsilon_K = v_t^3/L
\end{equation}

\noindent where $v_t$ is the rms turbulent velocity fluctuation at a given radius, 
and $L$ is a ``damping length''.  Dissipation calculated according to this expression is shown
in Figure~\ref{fig:oxygen-analysis} by the thin line and compares remarkably well
to the inferred damping rate, strongly supporting our ansatz about the nature of the
dissipation.  The damping length $L$, which represents the largest
eddy in the system, is comparable to the depth of the convective shell. 
Solar surface convection simulations (R.Stein, private communication) also
show this type of dissipation, though the dissipation length $L$ is about
four pressure scale heights and not the entire depth of the convection zone, but
still significantly larger than a pressure scale height.

\par In \cite{arnett2008} we consider the implications that this form of dissipation
has for the mixing length theory (MLT) of convection. In particular, we show that
if the dissipation length $L$ scales with the depth of the convection zone,
the near balance between dissipation and buoyancy driving,

\begin{equation}
  \avg{\rho' g v_t} = \varepsilon_K \approx v_t^3/L
\end{equation}

\noindent implies that the mixing length parameter $\alpha$ is a function of the
depth of the convection zone,

\begin{equation}
  \alpha^2 \propto L/H_p.
\end{equation}

\noindent While this result depends on various measured properties of the specific
flow at hand (such as the correlation coefficient between temperature and
velocity fluctuations, $\alpha_E =\langle T' v'\rangle/(T'_{\rm rms} v'_{\rm rms})$)
comparing a wide range of diverse simulations suggest some universality
(e.g., $\alpha_E \approx 0.7$ for a both oxygen shell burning, solar surface, 
and ideal box convection simulations; see \cite{meakin2007c,arnett2008}). 
That $\alpha$ is not a universal constant appears to be a robust result.

\subsection{Entrainment and Buoyancy Flux}

\par The interaction of a turbulent convective region with a bounding
stably stratified layer is a long standing challenge to stellar interior
modelers.  Various phenomenological models have been formulated to treat
the mixing that takes place at this interface, but can generally be classified
as either (1) a ballistic picture in which eddies penetrate the stable layers
until buoyancy breaking halts their motions (\cite{zahn1991}); (2) a
diffusive type process operating within the stable layers which
mixes material from the convection zone into the surroundings; or
(3) an instantaneous mixing in a region of a fixed, parametrized size.
While these prescriptions for mixing have been able to solve
various astrophysical quandaries (such as cluster color-magnitude diagram
fitting), they are not based on robust, self-consistent physical models
and contain parameters which are not grounded in more basic physical
considerations and must be calibrated.  The universality of these parameters
is therefore under question, and like the $\alpha$ in MLT are likely not
universal constants.

\par
\begin{figure}
  \includegraphics[scale=0.38]{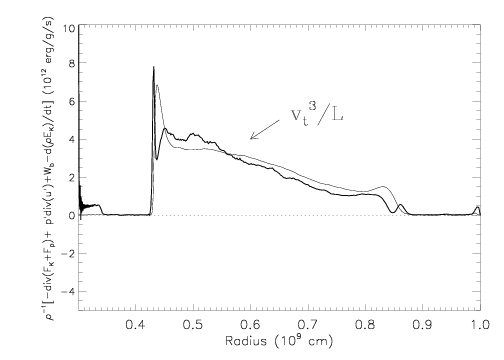}%{oburn_dissipation.ps}
  \includegraphics[scale=0.38]{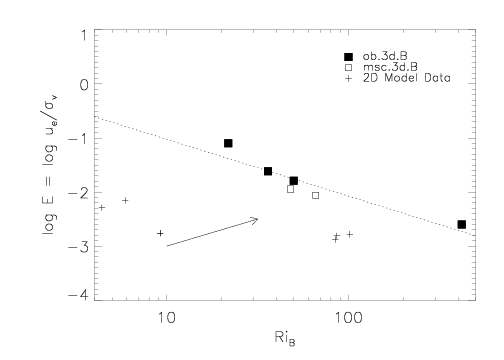}%{entrainment_diagram.ps}
  \caption{Derived flow properties for a numerical simulation of turbulent oxygen 
    shell burning in a 23\msol stellar model.
    (left) The inferred turbulent kinetic energy dissipation rate (thick line)
    is compared to a dissipation rate calculated using the local rms turbulent velocity 
    $v_t$ according to $\varepsilon_K = v_t^3/L$ with ``damping length'' $L$.
    The rms turublence velocity is based on the velocity variance across the horizontal
    plane and two convective turnover times for a fixed radius.  The damping length
    represents the largest eddy in the simulation, which is found to be comparable to the 
    depth of the convection zone in this model.
    (right) The normalized entrainment rate (boundary layer migration rate
    in units of the rms turbulence velocity at the boundary) is plotted against
    the bulk Richardson number, $Ri_B$. The dashed line shows the best-fit power
    law to the 3D data. The 2D entrainment rates fall everywhere below the
    3D trend due to the incompatible turbulence mixing properties between
    2D and 3D flow (see \cite{meakin2007c}).
    \label{fig:oxygen-analysis}
  }
\end{figure}

 A more detailed and rigorous analysis of the dynamics taking place at a 
turbulent boundary layer has been considered by both the geophysical and 
laboratory fluid dynamics communities, and much progress has been made
in elucidating fundamental processes which mediate the mixing rates at
these boundaries. One of the primary indicators of boundary layer dynamics
is the bulk Richardson number (\cite{fernando1991}),

\begin{equation}
  Ri_{B} = \frac{b L}{v_t^2}
\end{equation}

\noindent for buoyancy jump $b$, outer scale $L$, and rms turbulent
velocity $v_t$.  For large values of $Ri_{B}$, the boundary layers
are strongly stratified compared to the strength of the turbulence and
mixing proceeds slowly and the boundary remains relatively undistorted.
Boundaries with small values of $Ri_B$ are strongly distorted by the
turbulence, which is attended by more rapid mixing rates.  Entrainment rates,
defined as a boundary layer migration speed $u$ 
normalized by the rms turbulent velocity scale, is often found to be
well characterized by a simple power law dependence ``entrainment law'',

\begin{equation}
  E = \frac{u}{v_t} = A Ri_{B}^{-n}
\end{equation}

\noindent where A and n are constants fitted to experimental and 
simulation data. These observations are connected to the underlying
hydrodynamic processes through the buoyancy evolution of the boundary
layers (see e.g., \S7 in \cite{meakin2007c}).  An interesting
result is that this same power law dependence also holds for
the astrophysical convection simulations analyzed in \cite{meakin2007c}
(see right panel in Figure~\ref{fig:oxygen-analysis}).

\par
\begin{figure}
  \includegraphics[scale=0.38]{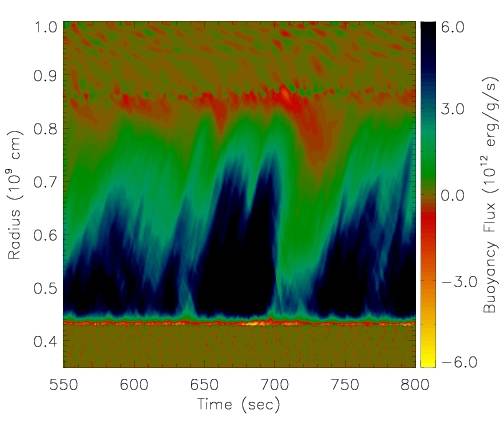}%{buoyancy_work_spacetime.ps}
  \includegraphics[scale=0.38]{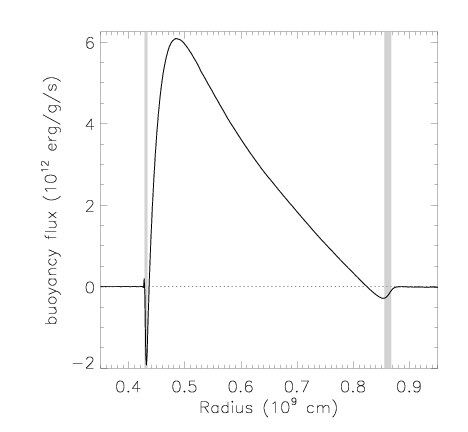}%{buoyancy_work_havg.ps}
  \caption{Buoyancy flux.  (left) Time-series diagrams and (right) time-averaged 
    radial profiles are shown for 3D oxygen shell burning model described
    in \cite{meakin2007c}.
    \label{fig:buoyancy-work}
  }
\end{figure}

 The evolution of buoyancy is related to the ``buoyancy flux'' through,

\begin{equation}
  \partial_t b = -\nabla(q)
\end{equation}

\noindent where $q = \rho' v' g/\rho_0$ which is related to the buoyancy work term 
in the Reynolds averaged KE equation above by $q = \rho_0\mathbf{W_b}$. 
This conservation law for buoyancy describes the exchange between the
kinetic energy in turbulence and the potential energy of stratification.
A fundamental theory of mixing at convective boundaries will model 
these terms (some progress is being made; see \cite{fernando1997,mcgrath1997}).
While the time and horizontally averaged profiles of the buoyancy flux is
smooth, a spatio-temporal decomposition reveals that the smooth profile
arises from a highly dynamic underlying behavior (Figure~\ref{fig:buoyancy-work}).

\begin{acknowledgments}
  I would like to thank the IAU for a travel grant that made my attendance at
  this meeting possible.
  This work is supported in part at the University of Arizona by 
  the National Science Foundation under Grant 0708871
  and by NASA under Grant NNX08AH19G.  This work was also supported in part
  at the University of Chicago by the National Science Foundation under Grant
  PHY 02-16783 for the Frontier Center ``Joint Institute for Nuclear Astrophysics'' (JINA).
\end{acknowledgments}

%\begin{discussion}
%  Some words were said.
%\end{discussion}

\end{document}